\documentstyle[11pt,adassconf]{article}  

\begin{document}

\paperID{P1-40}

\title{An Integrated Procedure for Tree-Nbody Simulations: FLY and AstroMD}

\author{U. Becciani, V. Antonuccio-Delogu}
\affil{Osservatorio Astrofisico di Catania, Catania - Italy}
\author{F. Buonomo,  C. Gheller}
\affil{Cineca,Casalecchio di Reno (BO) - Italy }

\contact{Ugo Becciani}
\email{ube@sunct.ct.astro.it}

\paindex{U. Becciani}
\aindex{V. Antonuccio-Delogu, F. Buonomo, C. Gheller }

\keywords{astronomy: parallel computing, data analysis, visualization}

\begin{abstract}          

We present a new code allowing to evolve three-dimensional self-gravitating 
collisionless systems with a large number of particles $N \geq 10^7$.  
{\bf FLY} (Fast Level-based N-bodY code) is a fully 
parallel code based on a tree  algorithm. It adopts periodic boundary 
conditions implemented by means of the Ewald summation technique.
FLY is based on the one-side communication paradigm to share data among the 
processors that access remote private data avoiding any kind of synchronism. 
The code was originally developed on a CRAY T3E system 
using the {\it SHMEM} library and it was ported on SGI ORIGIN 2000 and on 
IBM SP (on the latter making use of the {\it LAPI} library).
 \htmladdnormallinkfoot{FLY}{http://www.ct.astro.it/fly/} version 1.1 is an open 
 source {\it freely available code}.\\
FLY data output can be analysed with AstroMD, an analysis and visualization 
tool specifically designed to deal with the visualization and  analysis of 
astrophysical data. 
AstroMD can manage different physical quantities. It can find out structures 
without well defined shape or symmetries, and perform quantitative calculations
 on  selected regions. \htmladdnormallinkfoot{AstroMD}{http://www.cineca.it/astromd} is a 
{\it freely available} code.

\end{abstract}

\section{Introduction}
FLY is the tree N-Body code we design, develop and use to run very big 
simulations of the Large Scale Structure of the universe using parallel systems 
MPP and SMP. 
FLY uses the Leapfrog numerical integration scheme for performance reasons,  
and incorporates  fully periodic boundary conditions using the Ewald method.
The I/O data format is integrated with  the AstroMD package.\\
AstroMD is an analysis and visualization tool specifically designed 
to deal with the visualization and  analysis of astrophysical  data. 
AstroMD can find structures having a not well defined shape or symmetries, 
and  performs quantitative calculations on a selected region or structure.
AstroMD makes use of Virtual Reality techniques which are particularly effective
 for understanding the three dimensional distribution of the fields, their 
 geometry, topology and specific patterns. The display of 
data gives the illusion of a surrounding medium into which the user is 
immersed. The result is that the user has the impression of travelling through 
a computer-based multi-dimensional model which could be  directly 
hand-manipulated.

\section{FLY code}

The FLY code, written in Fortran 90 and C languages, uses the one-side 
communication paradigm: it has been developed on the CRAY T3E using the SHMEM 
library.
FLY is based on the following main characteristics. It adopts a simple
 domain decomposition, a grouping strategy and a data buffering that allows us 
 to minimize data communication.

\subsection{Domain decomposition}
FLY does not split the domain with orthogonal planes, but the domain 
decomposition is done by assigning an equal number of particles to each 
processor. The input data particles is a sorted file containing the fields of 
position and velocity, so that particles with a near tag number are also near 
in the physical space, and the arrays containing the tree properties are 
distributed using a fine grain data distribution.

\subsection{Grouping}
During the tree walk procedure, FLY builds a single interaction list (IL) to 
be applied to all particles inside a grouping cell ($C_{group}$). This reduces 
the number of the tree accesses to build the IL. We consider a hypothetical 
particle we call {\it Virtual Body} (VB) placed in the center of mass of the 
$C_{group}$: the VB interaction list $IL_{VB}$ is formed by two parts:
\begin{equation}
IL_{VB} = IL_{far} + IL_{near}
\end{equation}
where $IL_{far}$ includes the elements more distant than a threshold parameter 
from VB  and $IL_{near}$ includes the elements near  VB. Using the two lists  
it is possible to compute the  force $F_p$ of each {\it p} particle in 
$C_{group}$ as the sum of two components:
\begin{equation}
F_p =F_{far} +F_{near}
\end{equation}
The  component  $F_{far}$ is computed only once for VB, and it is applied 
to all the {\it p} particles, while the  $F_{near}$ component is computed 
separately for each  particle. The size of the $C_{group}$, and  the tree-level 
where it can be considered, is constrained by 
the  maximum allowed value of the overall error of this method. In this sense 
the performance of FLY is a level-based code.

\subsection{Data Buffering}
The data buffer is managed with a policy of a simulated cache in the local RAM.
Every time the PE has to access a remote element, at first it looks for the 
local simulated cache and, if the element is not found, the PE executes the 
GET calls to down-load the remote element and stores it in the buffer. 
In a simulation with 16-million-particles clustered, with 32 PEs and 256 
Mbytes of local memory, without the use of the simulated cache, the PEs 
execute about $2.1 \cdot 10^{10}$ remote GETs. This value, using the data 
buffering, decreases at $1.6 \cdot 10^8$ remote GETs, with an enormous 
advantage in terms of scalability and performance.\\
Fig. 1 shows the code performance on CRAY T3E system, obtained by running  
simulations with 32 and 64 PEs, and FLY scalability, considering 
the case of 16,777,216 particles, where a speed-up factor of 118 is reached 
using 128 PEs.
\begin{figure}
\plottwo{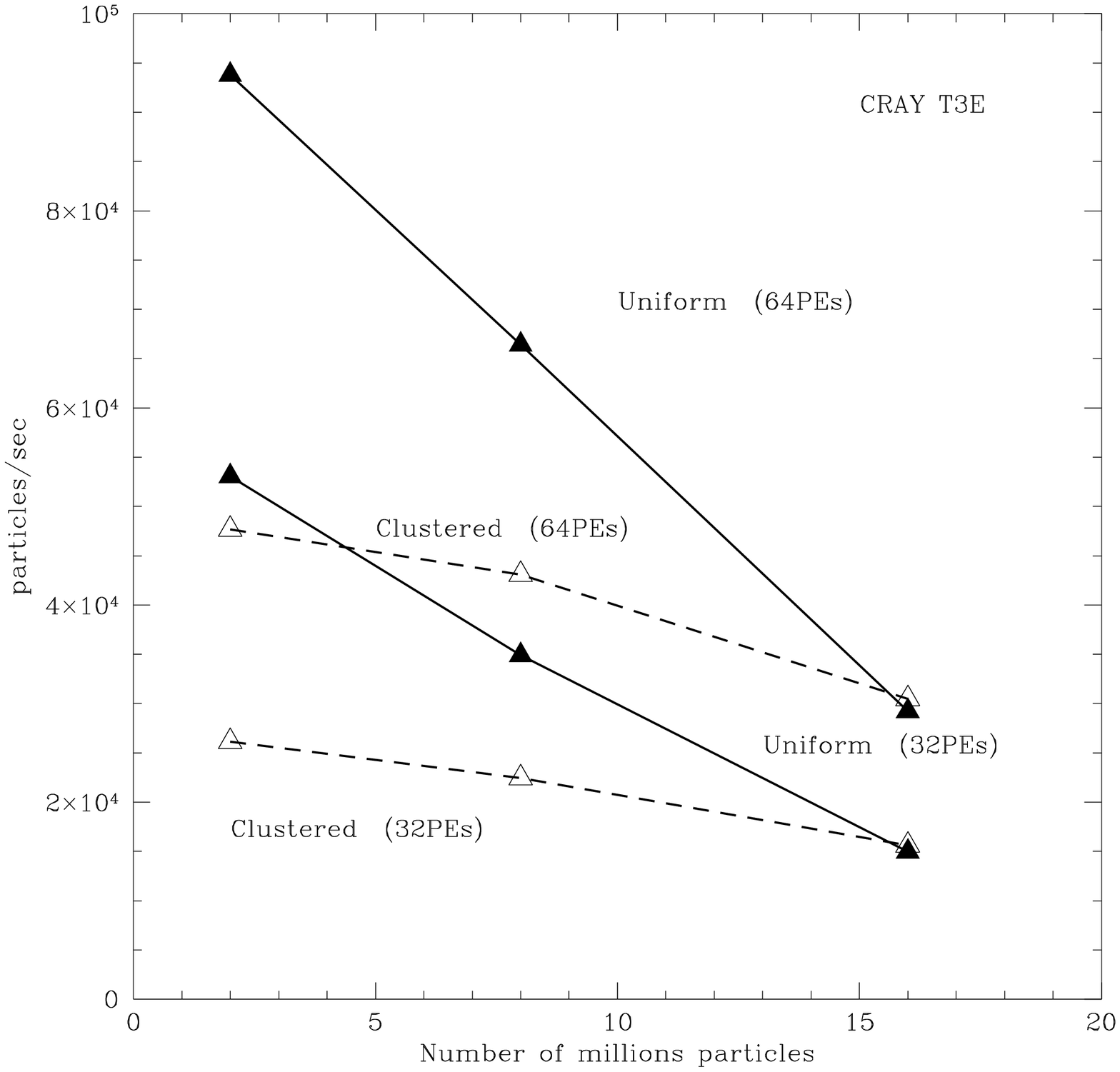}{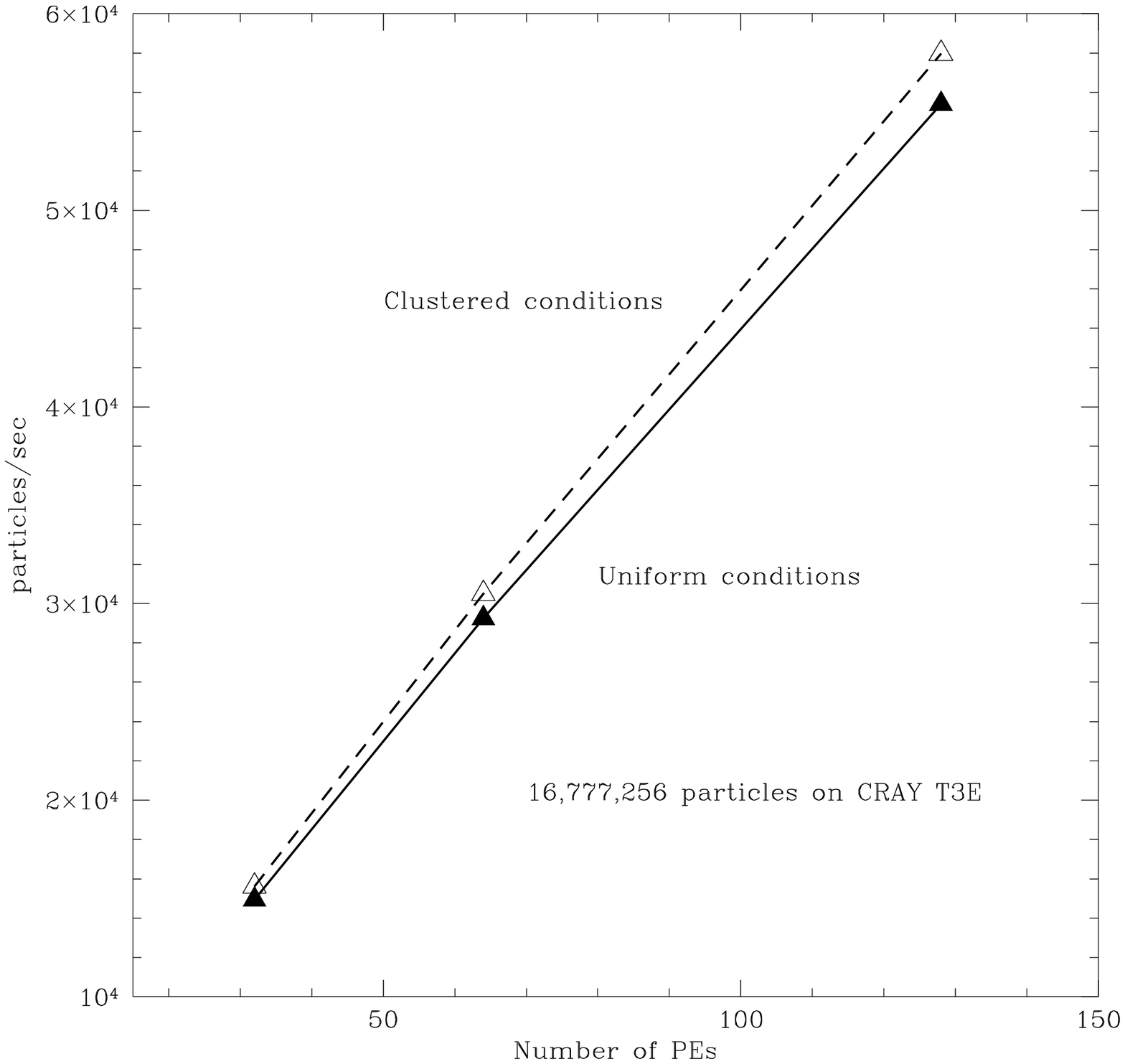}
\caption[h]{CRAY T3E/1200e. FLY particles/second using 32 and 64 PEs 
in uniform and clustered conditions, and scalability from  32 PEs 
to 128 PEs in a 16-million-particles simulation}
\end{figure}
\noindent The highest performance obtained in a clustered configuration is a 
positive effect of the grouping characteristic.
The obtained results show that FLY has a very good scalability and a very high 
performance and it can be used to run very big simulations.

\section{AstroMD package}

AstroMD is developed using the  Visualization Toolkit (VTK) by Kitware, a 
freely available software portable on several platforms which range from the 
PC to the most powerful visualization systems, with a good scalability. Data are 
visualized with respect to a box which can describe the whole computational mesh 
or a sub-mesh. AstroMD can find structures having a not well defined shape or 
symmetries, and performs quantitative calculations on a selected region or 
structure. The user can choose the sample of the loaded particle type 
(i.e. stars, dark matter and gas particles), the size of the visualization box 
and the starting time from which he wants to show the evolution of the simulation. 
The {\it Density} entries control the visualization of the iso-surfaces. These are 
calculated on a grid whose resolution can be user-selectable. To allow the user 
to investigate with more accuracy a subset  of the visualized system it is possible 
to use a cubic sampler. If the sampler is selected ({\it Show/Hide} menu), visualized in 
the scene and enabled ({\it Sampler} menu), all the computations are performed only 
inside the region of the sampler (Fig. 2).
\begin{figure} 
\plottwo{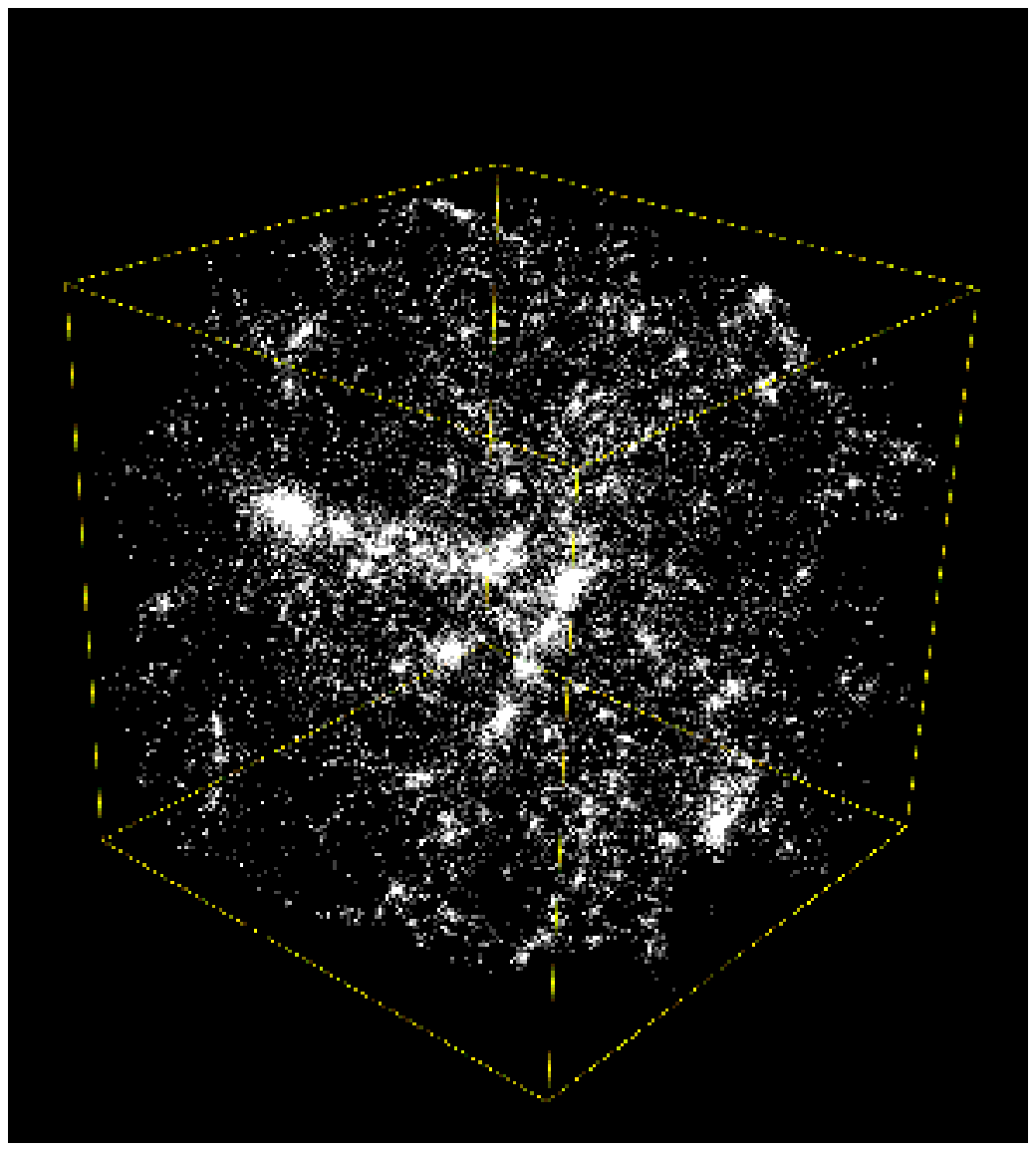}{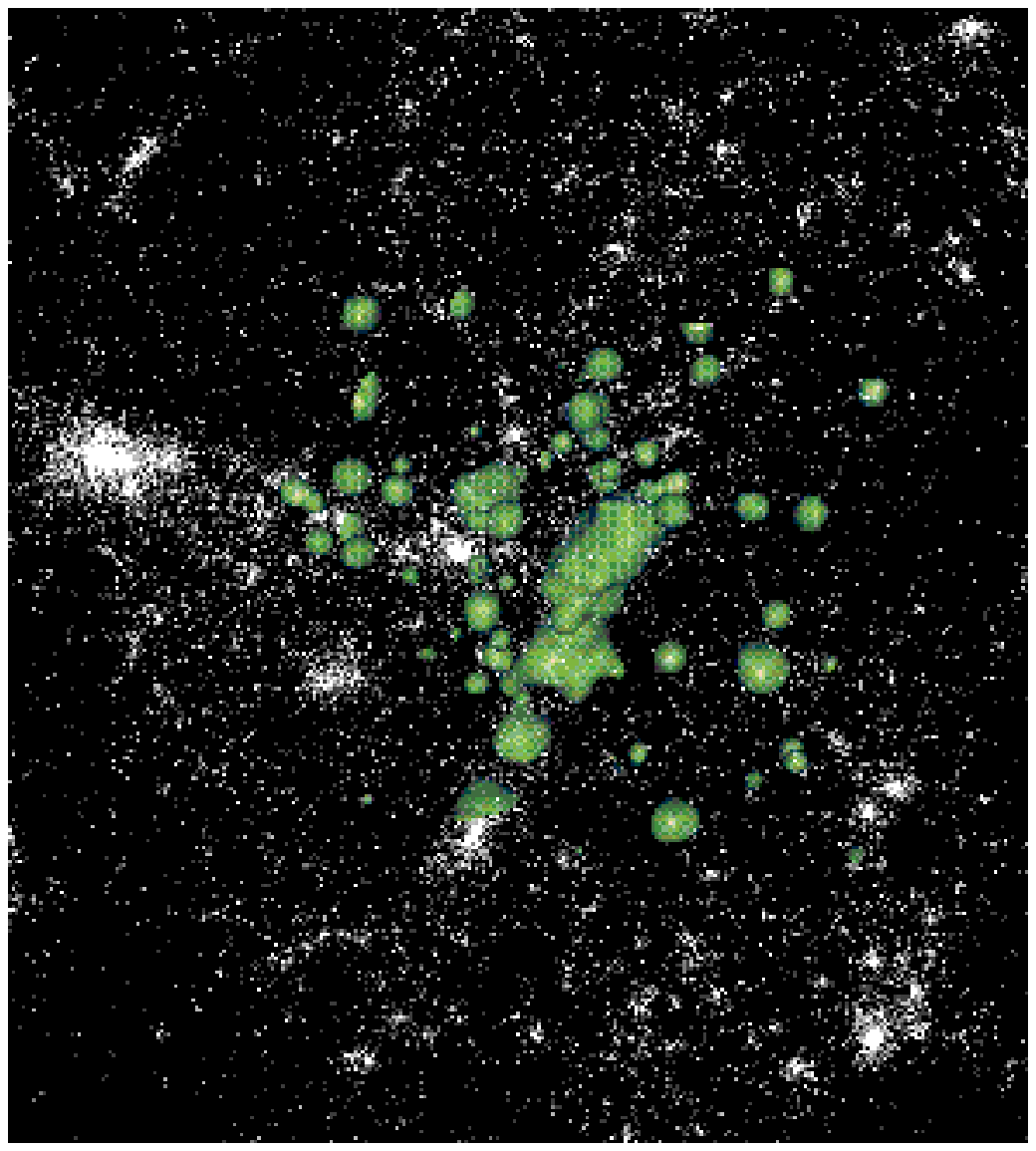}
\caption[h]{Typical structure of voids and filaments 
and Iso-surfaces, in a sub-region at the end of a simulation.}
\end{figure}
\noindent AstroMD can also show the evolution in time 
of the simulated system over all the interval of time for which the data are 
available, performing interpolations at intermediate frames. During the evolution 
the updated time of evolution is displayed  in the {\it Time} entry of the 
{\it Cloud} section.
Finally snapshots of the images displayed can be created using the button 
{\it Take it} in the {\it Screenshot} menu.
AstroMD is developed by the VISIT (Visual Information Technology) laboratory at 
CINECA (Casalecchio di Reno - Bologna) in collaboration with the 
Astrophysical Observatory of Catania.


\begin{references}

\reference Antonuccio-Delogu, V., Becciani, U.,  Pagliaro, A., Van Kampen, E.,
Colafrancesco, S., German\'a, A. and Gambera, M.\ 2000,  submitted to \mnras
\reference Barnes, J.E. and Hut, P.\ 1986, Nature, 324, 446\reference Schroeder, W., Ken, M. , Lorensen, B.\ 1999, "The Visualization Toolkit",  Prentice Hall
\reference Becciani, U., Antonuccio-Delogu, V.\ 2000, J. Comput. Phys., 163, 118 
\reference Earnshaw, R.A., Watson, D.\ 1993, "Animation and Scientific Visualization", Academic Press ltd. 
\reference Qinn, T., Stadel, C. TIPSY http://www-hpcc.astro.washington.edu/tools/TIPSY
\reference Salmon, J. and Warren, M.S.\ 1997, Proc.  of the Eight Conf. on Parallel Processing for Scientific Computing, SIAM 1997


\end{references}
\end{document}